\documentclass[11pt,onecolumn,amssymb,nofootinbib]{revtex4}
\usepackage{amsmath, amsthm, amscd, amssymb}
\usepackage{graphicx, braket}
\usepackage{bm}
\usepackage{bbm}
\usepackage{epstopdf}

\begin{document}

\title{\bf  Chen Ning Yang Retrospective  } \bigskip

\author{Stephen L. Adler}
\email{adler@ias.edu} \affiliation{Institute for Advanced Study,
Einstein Drive, Princeton, NJ 08540, USA.}

\begin{abstract}

Chen-Ning Yang made important contributions to the theory of solvable models in statistical mechanics, including generalizations of the Bethe Ansatz, magnetization in the Ising model, the Lee-Yang circle theorem, and the Yang-Baxter equation.  Most famously, Yang  made transformative contributions to  the current Standard Model of elementary particle interactions.  The proposal of Yang and T. D. Lee, that left-right symetry (parity) is violated in weak particle decays, established that the primary currents involved in weak interactions are left handed.  The work of Yang  and R. L.  Mills gave a framework for  force carriers coupling to these currents that are non-Abelian generalizations of the electromagnetic photon, which unlike the electrically neutral photon,  carry  ``charges'' to which they self-couple . Two decades of work by others on quantization and mass-generation mechanisms then culminated in the Standard Model.

\end{abstract}

\bigskip\bigskip\bigskip

\vfill\eject

\maketitle
Chen Ning (Frank) Yang had a remarkable life and career, spanning two centuries and embracing two continents and cultures.  He was born in Hefei, China in 1922, and despite having suffered a ``mild heart attack'' (II-239) in 1997 while at Stony Brook, and having bravely decided on quadruple bypass surgery, lived a long and productive life and died at the age of 103 in Beijing, China.  Yang made significant contributions to the field of Statistical Mechanics, and transformative contributions to the field of Elementary Particle physics.

A crowning achievement of physics in the decades since 1948 was the creation of the Standard Model of particle physics.  Many outstanding physicists contributed,  but in SLA's personal Pantheon of creators, two people, Frank Yang and Murray Gell-Mann, stand alone at the top.  In 2010  a conference in honor of Gell-Mann's 80th birthday was held in Singapore, and Yang came from China for the occasion.  An indelible image is Yang and Gell-Mann sitting conversing  together at a table, along with their ladies -- the two people who contributed the most to getting the field of High Energy Physics to where it stands now.  This judgement in hindsight was foreseen years before by Fermi:  ``In the fall of 1954, Murray was a visiting professor at Columbia.  He and I flew to Chicago to visit Fermi in the Billings Hospital....At the end of our visit when Murray and I got up and walked to the door, Fermi said to us from behind: `Now I have to leave physics to you guys.'''. (II-295)

Recounting his early education, Yang writes  ``In 1929 when I was seven years old, my father became a professor of mathematics at Tsinghua University in. ..(now Beijing), and we lived in a house on the campus for eight  years.'' (II-239)  These were idyllic years. Then, ``The War of Resistance Against Japan started in July 1937. We first moved back to Heifei, and then after the Japanese reached Nanjing, took a long journey... reaching finally Kunming in March 1938''. (II-230).  After a few months in eleventh grade, Yang ``skipping one year, entered the Southwest Associated University...in the fall of  that  year.''(II-230)

The National Southwest Associated University  (SAU) was formed by the wartime incorporation in Kunming of National Peking University, National Tsihghua University, and National Nankai University. Kunming, at an altitude of 6,234 feet on the Yunnan-Guizhou Plateau, was during World War II  a Chinese military center, transport terminus for the Burma Road, and home of  the Flying Tigers.  It was subject to constant air raids: ``On September 30, 1940,  the house that my family rented in Kunming received a direct hit, reducing most of our meager possessions to rubble. Fortunately, every member of the family was in some shelter and no one was wounded.'' (I-3) Moreover ``...there was the constant threat of inflation.  My father was a professor at the Southwest Associated University, and his savings were totally wiped out..... To feed and clothe a family of seven, my mother, a woman of great will power and self-discipline, toiled from dawn to night, year after year, with calm dignity.  The family survived the war intact --lean, very lean, but healthy.'' (I-4)

``I had not studied high school physics, so to prepare for the entrance examination [to SAU] I borrowed a copy of a standard high school physics textbook and read it through in several weeks...I concluded that physics was the subject that I liked ...[and] when I registered .. I enrolled in the Department of Physics.'' (II-308)

``In retrospect, my taste in physics was largely formed during the six years (1938-1944) I spent at'' SAU. (I-4)  Professor  T. Y. Wu at SAU suggested reading through which  ``I was..introduced to group theory in physics'',  which ultimately comprised ``approximately two thirds of my research efforts.'' (II-309)  Professor J. S. Wang at SAU ``specialty was statistical mechanics and he guided me into this area of research.  Approximately one third of my later research work was in this area.'' (II-309).  Yang profited from long discussions with fellow students.  ``...we argued endlessly about physics.'' Discussing the subtle topic of measurements  in quantum mechanics, ``our arguments started at  tea, lasted throughout the evening, and continued back in our [dorm] room....After the electric lights were turned off and we were all in bed, the arguments did not stop.'' (II-311)

 ``...my attitude toward what to like in physics was largely formed in those years in Kunming....It was in those years that I learned to admire the work of Einstein, Dirac, and Fermi,... who share the ability to extract the fundamentals of a physical concept, a theoretical structure, or a physical phenomenon and to zero in on the essentials....In contrast, I did not resonate with the style of Heisenberg.'' (I-4,5)  Yang admired Fermi both for his physics and also for his character.  ``He did not practice one-upmanship.  He exemplified, I always believe, the perfect Confucian gentleman.'' (II-246).

To be nearer to Fermi,   ``A few weeks after the end of World War II in 1945 I flew on a DC3 to Calcutta, where I had to wait for a berth on ships bound for the U. S. for several months.'' (II-312). ``In January 1946    I enrolled as a graduate student in the Physics Department of the University of Chicago. My aim was to write an experimental Ph. D. thesis with...Fermi.'' (II-312).  But Fermi's ``laboratory ...  was in the Argonne National Laboratory which was not open to me.'' (II-313)  So on Fermi's recommendation Yang worked on theory with Teller and experiment with Allison.  His attempt to be an experimentalist under Allison at Chicago proved abortive.  ``I was well received by my fellow graduate students in the laboratory, since I  occasionally could help them with theory.  But they told jokes about me.  The one that Allison liked especially was, ``Where there is a bang, there is Yang.'' (I-6) Eventually  Teller swooped in to the rescue, and under his aegis a short theory paper that Yang wrote applying group theory to angular distributions of products of nuclear reactions became Yang's PhD thesis.

Striking out on his own,  Yang wrote a celebrated paper applying group theory to the newly discovered pi meson decay to two photons. Teller had given a too simple argument that ``when challenged was demolished.  I later thought about the matter and the next day worked out the correct selection rules.'' (I-8) This was Yang's ``second published work on symmetry principles.  It anchored my interest in the field.'' (I-8) ``These  two papers established me as one of the foremost theorist on the use of group theory in analyzing symmetry properties.'' (II-315).

In 1948, when Yang received his PhD, ``the hottest theoretical area of research was renormalization theory.  The three theory professors in Chicago ... were not working in this field. Therefore after one year  1948-1949 I moved to the Institute for Advanced Study  (IAS) in Princeton....Fermi told me it was good to spend some time at the IAS, but the work there was too academic.  It was a bit like a medieval cloister, he said, and was not a place for [a] long stay....I was of course totally in agreement... But because of the convenience of dating Miss [Chih Li] Tu (later my wife) in New York City, I did not return to Chicago after one year and instead remained at the IAS for altogether seventeen years, 1949-1966.'' (II-315) Despite Fermi's well-intentioned advice, the IAS evidently suited Yang's temperament, because the majority of his most famous work was done while he was a Professor there.

Yang's new wife Chih Li Tu was the  daughter of the Nationalist general Du Yuming, who was captured by Mao's army and served ten years in prison before being pardoned by Mao and residing in Beijing.  She had been in a class Yang taught at Kunming high school, but Yang did not know her well then. Yang ran into Chih Li (then studying in New York)  by chance four years later when having dinner at the Tea Garden, the only Chinese restaurant in Princeton.   ``Eight months later'', ``Chih Li and I were married in Princeton on August 26, 1950). (I-11, II-233)

Yang was originally appointed to the IAS as a long-term Member, but in a few years was promoted to the permanent Faculty. Settling in Princeton was not without glitches.  ``In late 1954, my wife and I paid a couple of hundred dollars as a deposit for a home in a new development near Princeton.  A few weeks later we were told by the developer that he had to return our deposit because he was afraid that our being Chinese might affect his sales. We were furious and talked to a lawyer.  He advised us not to sue, since in his opinion we had little chance of winning.'' (I-57) Restrictive covenants on residential housing (stipulating Aryan only, etc.) were not uncommon in Princeton until a community-wide open housing drive in the 1960's, and only in 2021 were  banned by state legislation in New Jersey.

Despite this unpleasant experience, Yang  wrote  ``Yes,  there were things that held me back.  Yet I knew that America had been most generous to me.  I had come very well equipped, but America had allowed me the opportunity to develop my potential.  I knew  there was no country in the world that was as generous to immigrants.   I also realized that.. my roots here were deepening..''. (I-57)  In 1961 Yang watched Robert Frost read his poem ``The Gift Outright'' at the Kennedy inauguration, and ``something seemed to go directly to my heart....It was to play a part in my decision to apply for U.S. citizenship.'' (I-57)

At the IAS, Yang returned to an earlier interest in the statistical mechanics of the two dimensional Ising model. He realized that work of Onsager and Kaufman gave more information than originally thought, allowing a calculation of the spontaneous magnetization.  ``I was thus let to a long calculation, the longest of my career.  Full of local, tactical tricks, the calculation proceeded by twists and turns...Finally, after about six months of work off and on ...I was staring at the amazingly simple final result.''  That bravura result ``was in June, 1951, about a week before my first child, Franklin, was born.'' (I-12)

Also at the IAS, Yang resumed earlier work with T. D. Lee, another immigrant from China.  ``Lee had enrolled at the University of Chicago in the fall of 1946.  Although we had probably met earlier in China, it was in Chicago that we really got acquainted.  I found him to be exceptionally bright and hard working.  We got along well and soon became close friends. Being older and several years ahead of him in my graduate studies, I tried to help him in every way.'' (I-7)  Thus began a very fruitful multi-year collaboration, but in the wording of Yang above one can already see the seeds of the seniority issue that eventually led to their bitter breakup.

 ``In the fall of 1951, T. D. Lee came to the Institute... and we resumed our collaboration.'' (I-14)  In statistical mechanics, they studied a simple model called a lattice gas, ending up, with the stunning result that ``the roots of the partition functions, which are polynomials in the fugacity, are all on the unit circle...The theorem, later called the unit circle theorem, became the main element ...to discuss the thermodynamics of a lattice gas.'' (I-15).

Over the years Yang continued to produce important work on statistical mechanics.  In collaboration with his younger brother Chen Ping Yang (a professor at Ohio State) he wrote in 1969 an important paper applying the Bethe Ansatz to the thermodynamics of a one-dimensional system of bosons with repulsive delta-function interactions. And in studies relating to the braid group acting on three strands, Yang formulated what became known as the Yang-Baxter equation, which contributed to the development of important integrable models.

Yang's most famous work, however, was in applications of symmetries to particle physics.  From reading starting in Kunming, Yang ``was very much impressed with the idea that charge conservation was related to the invariance of a theory under phase changes...I was even more impressed with the fact that gauge invariance {\it  determined} all electromagnetic interactions. When in Chicago, I tried to generalize this to isotopic spin interactions'' [rotations in charge space such as from proton to neutron]   but initial attempts ``led to a mess, and I had to give up'' (I-19).  However, the general idea became an ``obsession'' (I-19) to which Yang returned repeatedly. During a 1954 sabbatical from the IAS,``while at Brookhaven I returned once more to the idea of generalizing gauge invariance,'' this time working with  office-mate Robert L. Mills. (I-19)  They tried to eliminate the ``mess'' by adding polynomial terms to the relation between field strength and gauge potentials. ``We decided to first try a quadratic polynomial.  If that did not work we would try a cubic one...Fortunately, we rapidly found that if we [added a simple quadratic term] the subsequent calculation became {\it increasingly simple}.  Thus we knew that we had uncovered a great treasure!!!'' (II-319)  This was written up as classic papers, and Yang-Mills theory, or in technical terms non-Abelian gauge theory,  became a foundation for the subsequent unification of particle forces.  The questions of mass, emphasized in critique by Pauli (who had not published his own similar ideas), and of quantization, took two decades of work by a score of brilliant theorists to resolve.

In 1956 Yang, while on summer leave at Brookhaven, working with T. D. Lee,   made a comprehensive study of whether experiments to date had established parity (spatial reflection symmetry) as a valid symmetry of beta type weak particle decays.      This was motivated by the so-called theta-tau puzzle, the fact that particle(s) of the same mass decayed into even and odd parity final states of pions. ``The result was that, in all these processes, previous experiments did not yield any information about whether there was only [one type of] interaction or there were both types of interaction [i.e., mixtures of even and odd parity terms].  In other words, {\it all previous beta decay experiments were irrelevant as far as the question of parity conservation for beta decay was concerned}.'' (I-28).  The underlying reason was that ``terms proportional to [products of the two types] in the calculations must be 'pseudoscalars'.  Since all previous experiments did not measure a pseudoscalar, they had therefore no bearing on parity conservation in beta decay.'' (I-28,29).

Experiments measuring pseudoscalars that involved particle spins were soon done, and gave the sensational result that parity is not conserved in the weak interactions.  This was front page news in the NY Times, and two years later Lee and Yang shared the 1957 Nobel Prize for their incisive analysis.  They were the first Chinese to win a Nobel Prize. Yang's parents were living then, and Yang was in regular touch with them, but China was still very isolated and his parents were not recorded as present at the ceremony in Stockholm.

Other important symmetry related work included a 1957 paper with T. D. Lee and Reinhard Oehme studying parity (P), charge conjugation (C), and time reversal (T) symmetries, which included an early analysis of the neutral K meson--anti K meson system.  In 1964, after violation of the product symmetry charge conjugation times parity (CP) was discovered experimentally in this system, Yang in collaboration with his younger protege T. T. Wu decided to avoid the rush towards premature speculations as to its origin.  Instead, ``With our tendency toward restraint, Wu and I decided to make a phenomenological analysis of kaon-antikaon decay....It provided the framework within which subsequent experiments... were analyzed.'' (I-58,59) The collaboration with T. T. Wu continued, and a decade later on Yang and Wu wrote important papers on formal aspects of gauge theories and their relationship to the mathematics of fiber bundles.

Shortly after the CP discoveries, Oppenheimer decided to retire as IAS Director, and told Yang that ``he would propose to the Trustees that I be appointed his successor...I thought it over and some days later wrote him...It is quite uncertain that I shall make a good Director, while it is quite certain that I shall not enjoy the life of a Director.'' (I-60) Likely Yang had in mind in writing this the ongoing  acrimony between some of the Mathematics Faculty and the Director.    Not that Yang was incapable of the occasional sharp remark.  In his roast of Gell-Mann at Murray's 80th birthday conference in Singapore mentioned above, Yang said ``Anyone who had had contact with Murray cannot fail to be impressed by his catholic interests in many things, by his knowledge,  by his humor, but also by his sometimes overbearing self-confidence.'' (II-295) In terms of having Murray as a colleague at the iAS this last tipped the scale.  ``In the late 1950s, at a physics faculty meeting at the Institute for Advanced Study, Oppenheimer said casually he was thinking of asking Murray to join the Institute.  At the next meeting, I said Murray is great, but if he comes to the Institute, I shall leave.  Oppenheimer never mentioned the subject again.'' (II-295).

 ``But destiny seemed to be arranging things to change my career anyway ... During 1964-1965, the legislature of the State of New York voted to establish five Einstein Professorships at universities within the State....John Toll,..., T. A. Pond, and M. Dresden... decided to approach me to accept an Einstein Professorship at Stony Brook that they hoped to bid for....Toll and Pond offered to have me head an Institute of Theoretical Physics to be built up over the next few years. '' (I-60)   The entire  Yang family visited  ``Stony Brook in the spring of 1965. We were put up at Sunwood, the University's guest house overlooking the Long Island Sound. The first evening we were there, the window of our room framed a spectacular sunset over the sound.  It captured our hearts ....Around the end of April, I accepted the Stony Brook offer and told Toll I would take up my post in 1966.''  (I-60)

``For my family, the move to Long Island was exciting. For  me the feeling was more complex.  I had spend seventeen years, 1949-1966, at the Institute for Advanced Study, from  age 27 to age 44.  I had been productive and happy there. I liked its unpretentious Georgian buildings and its peaceful, restrained atmosphere.  I liked its long, meandering walk to the little suspension bridge in the woods. It was a place out of this world. It was a place meant for contemplation, and it was populated by people who contemplated well.  The permanent faculty was first class.  The visitors were generally brilliant.  It was an ivory tower in the best sense of the term.''(I-64)

Asking whether he ``was making the right decision to leave the Institute...the answer was always the same. Yes, it was the right decision:  The ivory tower is not the world, and the challenge to help build a new university is exciting.'' (I-64). Despite his earlier reservations about the Directorship at IAS, Yang made an excellent Director at Stony Book, where he stayed for the 33 years 1966-1999.  He built up an outstanding group, did important research and he, as well as his family, enjoyed the change  of ambiance.

But this was not to be Yang's final move.  ``In the summers of 1960 and 1962, Father and Mother and I had reunions again in Geneva.....Father had, during his three trips to Geneva, especially during the last two, a sense of responsibility to convince me to go back to China .  This was ...partly due to....suggestions of the Chinese Government.  But is was also due to a desire in the great depths of his own soul.'' (II-236)  This sowed a seed that sprouted four  decades later, when Tsinghua University made Yang an offer and organized a symposium to celebrate his 80th birthday.  The timing was opportune. ``My heart bypass operation was successful, but Chih Li's health problems mutliplied, requiring several operations....After Chih Li passed away on October 19, 2003,  I moved to Beijing to Tsinghua University in December.'' (II-271)  At the Tsinghua 80th birthday conference, Yang remarked ``I am ...  blessed with the opportunity to launch an exciting new career helping this university to build a Center for Advanced Study.   The Tsinghua campus is where I had grown up.  My life has now completed a big circle.'' (II-255)

The move back to China also led to a big change in Yang's personal life.  On November 7, 2004 Yang announced his engagment to Miss Fan Weng, 52 years his junior, and they were married December 24, 2004.  She was Yang's constant companion for the final two decades of his life, ``God's benevolent last gift To give my old soul A joyous rejuvenating lift.'' (II-279).

Although Yang expressed deep appreciation for the ``many, many papers on gauge theories in the 1960s and 1970s'' (I-67), he remained aloof from these efforts that led to the current Standard Model. ``But I continue to believe that fundamental new ideas are still missing.  For example, the introduction of a field to break symmetry cannot be the final story, although it may be a good temporary development, perhaps not unlike Fermi's theory of beta decay.''  (II-67) The mystery of the Higgs field is still a deep topic for investigation by current and future high energy theory and experiment.

In the Preface to volume (II) of Yang's Selecta, he wrote that while the first volume  ``covered the years when my main interests were in physics research,  the present one covers later years when my interests  gradually shifted to history of physics.'' (II-v) Many of these articles are reprinted in II, and Yang's astute assessments of the great physicists who were his predecessors make fascinating reading.  After completion of II, the published record ceases, like a footpath that has petered out in a dense forest. Yang's summation  ``On Reaching Age Ninety'' suffices:

''Mine has been, A promising life, fully fulfilled, A dedicated life, with purpose and principle, A happy life with no remorse or resentment, and  long life ....Traversed in deep gratitude.'' (II-342)

\bigskip
\bigskip
\bigskip

\vfill\eject

{\bf  Bibliographic note}

 The online searchable Wikipedia article on Yang, and the New York Times obituary, are standard sources. Fascinating information not in these is contained in Yang's two volume Selecta, from which we quote extensively,  followed by the volume (I or II)  and page in parentheses.

(I) Chen Ning Yang, ``Selected Papers 1945-1980 with Commentary, W. H. Freeman and Company, San Fransisco (1983)

(II) Chen Ning Yang, ``Selected Papers II With Commentaries'', World Scientific, Singapore (2013)

\end{document}